\documentclass[10pt,twocolumn,floatfix,superscriptaddress,preprintnumbers,amsmath,amssymb,prb]{revtex4-2}
\usepackage{graphicx}
\usepackage{dcolumn}
\usepackage{bm}
\usepackage{amssymb,amsmath}
\usepackage{float}
\usepackage{color}
\usepackage{ulem}

\begin{document}

\preprint{preprint(\today)}

\title{Ti$_4$Ir$_2$O a time-reversal-invariant fully gapped unconventional superconductor}

\author{Debarchan Das}
\affiliation{Laboratory for Muon Spin Spectroscopy, Paul Scherrer Institute, CH-5232, Villigen PSI, Switzerland}
\author{KeYuan Ma}
\affiliation{Max Planck Institute for Chemical Physics of Solids, 01187 Dresden, Germany}
\author{Jan Jaroszynski}
\affiliation{National High Magnetic Field Laboratory, Florida State University, Tallahassee, Florida 32310, USA}
\author{Vahid Sazgari}
\affiliation{Laboratory for Muon Spin Spectroscopy, Paul Scherrer Institute, CH-5232, Villigen PSI, Switzerland}
\author{Tomasz Klimczuk}
\affiliation{Faculty of Applied Physics and Mathematics, Gdansk University of Technology, ul. Narutowicza 11/12, Gda\'{n}sk 80–233, Poland}
\affiliation{Advanced Materials Centre, Gdansk University of Technology, ul. Narutowicza 11/12, Gda\'{n}sk 80–233, Poland}
\author{Fabian O. von Rohr}
\affiliation{Department of Quantum Matter Physics, University of Geneva, CH-1211 Geneva, Switzerland}
\author{Zurab~Guguchia}
\affiliation{Laboratory for Muon Spin Spectroscopy, Paul Scherrer Institute, CH-5232, Villigen PSI, Switzerland}

\begin{abstract}
Here we report muon spin rotation ($\mu$SR) experiments on the temperature and field dependence of the effective magnetic penetration depth $\lambda(T)$ in the $\eta$-carbide-type suboxide Ti$_4$Ir$_2$O, a superconductor with an considerably high upper critical field. Temperature dependence of $\lambda (T)$,  obtained from transverse-field (TF)-$\mu$SR measurements, is in perfect agreement with an isotropic fully gaped superconducting state. Furthermore, our ZF $\mu$SR results confirm that the time-reversal symmetry is preserved in the superconducting state. We find, however, a remarkably small ratio of $T_{\rm c}$/$\lambda^{-2}_{0}$ $\sim$1.22. This value is close to most unconventional superconductors, showing that a very small superfluid density is present in the superconducting state of Ti$_4$Ir$_2$O. The presented results will pave the way for further theoretical and experimental investigations to obtain a microscopic understanding of the origin of such a high upper critical field in an isotropic single gap superconducting system.

\end{abstract}


\maketitle


The quest for superconducting materials for groundbreaking applications continues to drive the condensed matter research community, leading to the discovery of an array of novel superconductors and improved superconducting properties. \cite{Zhou}. A lot of research effort has been put forward to advance novel superconducting materials with improved properties such as high transition temperature ($T_{{\rm c}}$), high value of critical current, and of upper critical field ($H_{\rm c2}$). Furthermore, high upper critical fields $H_{\rm c2}$ are of interest from a fundamental point of view, as these are commonly associated with unconventional, possibly even topological, superconductivity.\cite{Falson,Khim,vonRohr} High-upper critical fields are also associated with unusual field-induced superconducting states, such as e.g., the Fulde-Ferrell-Larkin-Ovchinnikov (FFLO) state, which is a distinct superconducting phase observed in spin-singlet superconductors.\cite{Fulde,Larkin,Clogston}

One recent approach to control the properties of superconductors has been the chemically precise filling of void position in the crystal structures using electron-donor atoms to chemically tune the electronic properties of these materials.\cite{Zhang,Balestra} Along this line, $\eta$-carbide-type oxides turn out to be a promising material platform since it results from filling void positions in the Ti$_2$Ni-type structure, with small non-metallic atoms such as oxygen, nitrogen, or carbon occupying the interstitial positions\cite{Mackay}.

$\eta$-Carbide family of compounds, encompasses a variety of materials which are important for the investigation of emergent quantum properties \cite{Ma, Ku, Waki, Ma2, Ma3}. Most importantly, some members of this family exhibit superconductivity \cite{Ma3, Ma2, Ruan}. One notable example is Ti$_4$Ir$_2$O which manifests superconducting properties with a critical temperature of $T_{\rm c}$ $\simeq$ 5.3~K along with a considerably high upper critical field of $\mu_0 H_{\rm c2}$ $\sim$16~T as revealed from bulk measurements\cite{Ma3}. Interestingly, the $H_{\rm c2}$ exceeds the weak-coupling Pauli limit ($\mu_0 H_{Pauli} \approx$ 9.8~T) which corresponds to the paramagnetic pair breaking effect governing the maximal $H_{\rm c2}$ in conventional superconductors \cite{Tinkham}. Recent DFT calculations revealed the presence of multiple bands crossing at the Fermi energy $E_{\rm F}$ signaling plausible multi-gap superconductivity in Ti$_4$Ir$_2$O. Moreover, there is a pronounced peak of density of states close to $E_{\rm F}$, arising from the weakly dispersing energy bands\cite{Ruan}. This aspect resembles a key aspect of the recently widely discussed flat-band kagome superconductors \cite{Kiesel,Ortiz, MielkeGuguchia,GuguchianNPJ}. Most recently, it was found that Ti$_4$Ir$_2$O maybe the first fully isotropic FFLO superconductor, in contrary to the previous candidates, which all display highly anisotropic layered structures.\cite{Lortz}

To gain insight into the superconducting properties of Ti$_4$Ir$_2$O, we conducted muon spin rotation/relaxation ($\mu$SR) experiments, a technique highly sensitive to superconducting gap structures in materials \cite{Sonier,Blundellbook, Hillierreview}. Using transverse field (TF) $\mu$SR in type-II superconductors, we can assess the temperature dependence of the magnetic penetration depth $\lambda$, crucial for understanding the superconducting gap structure. Additionally, zero-field (ZF) $\mu$SR  is effective for detecting very small internal magnetic fields, ideal for exploring spontaneous magnetic fields related to broken time-reversal symmetry in exotic superconductors \cite{GuguchiaRVS, LukeTRS, Hillier, Barker, RPSingh, ShangRebased, Biswas}.

In this letter, we report on the low density of copper pairs (dilute superfluid) and unusually high upper critical field larger than the Pauli limit in $\eta$-carbide-type suboxide Ti$_4$Ir$_2$O. We, furthermore, demonstrate the isotropic fully gap pairing and the preserved time-reversal symmetry in the SC state of this system. These results identify Ti$_4$Ir$_2$O as time-reversal-invariant fully gapped unconventional superconductor.

\begin{figure}[htb!]
\includegraphics[width=1.0\linewidth]{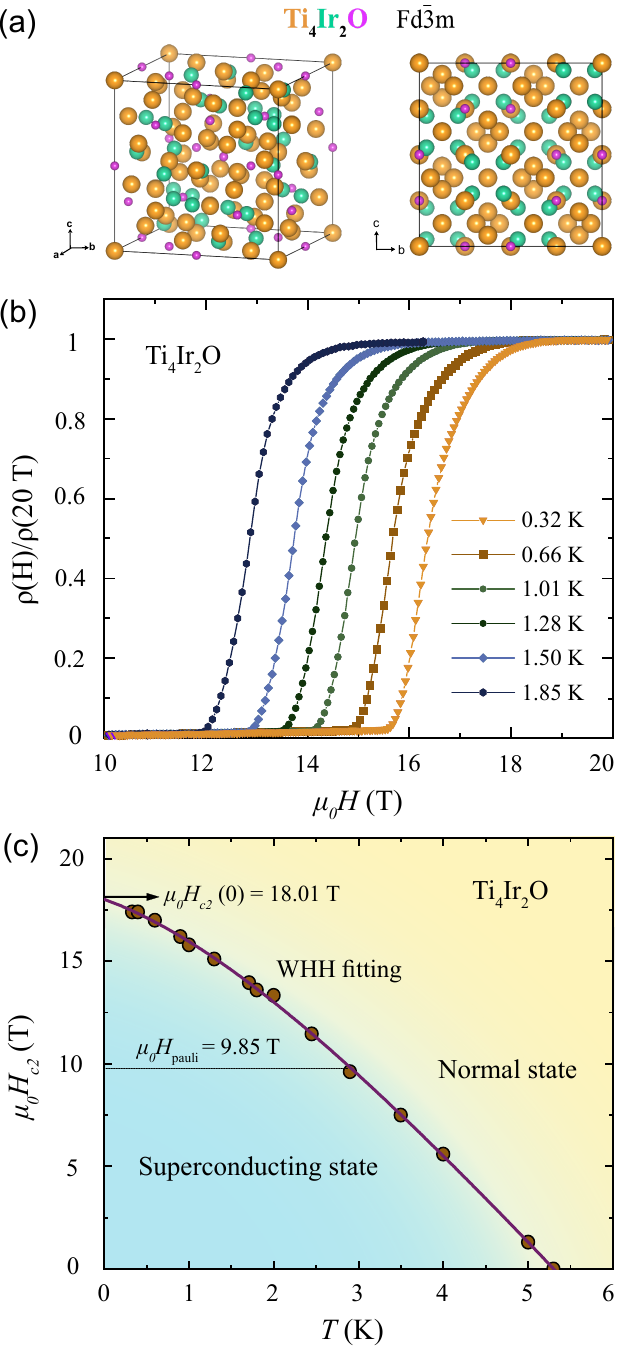}
\caption{(a) Schematic representation of two orientations for the crystal structure of Ti$_4$Ir$_2$O, (b) high magnetic field (10 - 20~T) transport measurement of Ti$_4$Ir$_2$O at low temperatures down to 300 mK (c) $\mu_0H_{\rm c2} (T)$ phase diagram of Ti$_4$Ir$_2$O determined from electrical transport measurements under external fields. The solid purple line represents the fitting of the experimental data with Werthamer-Helfand-Hohenberg (WHH) expression.}
\label{fig1}
\end{figure}


Polycrystalline Ti$_4$Ir$_2$O samples were synthesized from high-purity titanium (purity: 99.9\%, Alfa Aesar), iridium (purity: 99.99\%, Strem Chemicals), and titanium dioxide powders  (purity: 99.9\%, Sigma-Aldrich) via arc-melting and solid-state reaction. Initially, the reactants were mixed in stoichiometric ratios, formed into a pellet, and melted in an arc furnace under a purified argon atmosphere to ensure homogeneity. The resulting melt was then ground into fine powders, re-pelletized, and annealed in a sealed quartz tube at 1000 °C for 7 days under a partial argon atmosphere. The reaction was completed by cooling the quartz tube to room temperature through water quenching.

For magneto-transport measurements, a bar shaped sample was used with four 50~$\mu$m dia platinum wire leads spark-welded to the sample surface. The transport measurements were performed in the 35 T resistive magnet at the National High Magnetic Field Laboratory. The magnet was swept from 11.5 T to 20 T owing to the static 11.5 T background field generated by the outer superconducting coil of the magnet combination. The temperature under 2 Kelvin was ascertained using the vapor pressure of Helium-3, which remains unaffected by magnetic fields. This method contrasts with the use of Cernox and Ruthenium Oxide thermometers, which exhibit significant magnetoresistance. Transport measurements in lower fields were performed using a Physical Properties Measurements System (PPMS) from Quantum Design, equipped with a 9T superconducting magnet.

Transverse-field (TF) and zero-field (ZF) ${\mu}$SR experiments were carried out at $\pi$E1 beamline on DOLLY spectrometer at the Paul Scherrer Institute (Villigen, Switzerland). The field dependent TF-${\mu}$SR experiments at 1.6~K were were performed at $\pi$M3.2 beamline using GPS spectrometer. For ${\mu}$SR experiments, we used powdered sample which was pressed into a 7~mm pellet and then mounted on a Cu holder using GE varnish. This holder assembly was then mounted in the respective spectrometer cryostat. Both spectrometers are equipped with a standard veto setup \cite{Amato} providing a low-background ${\mu}$SR signal. All the TF experiments were performed after field-cooled-cooling the sample. The ${\mu}$SR time spectra were analyzed using the MUSRFIT software package~\cite{Suter}.


\begin{figure}[htb!]
\includegraphics[width=1.0\linewidth]{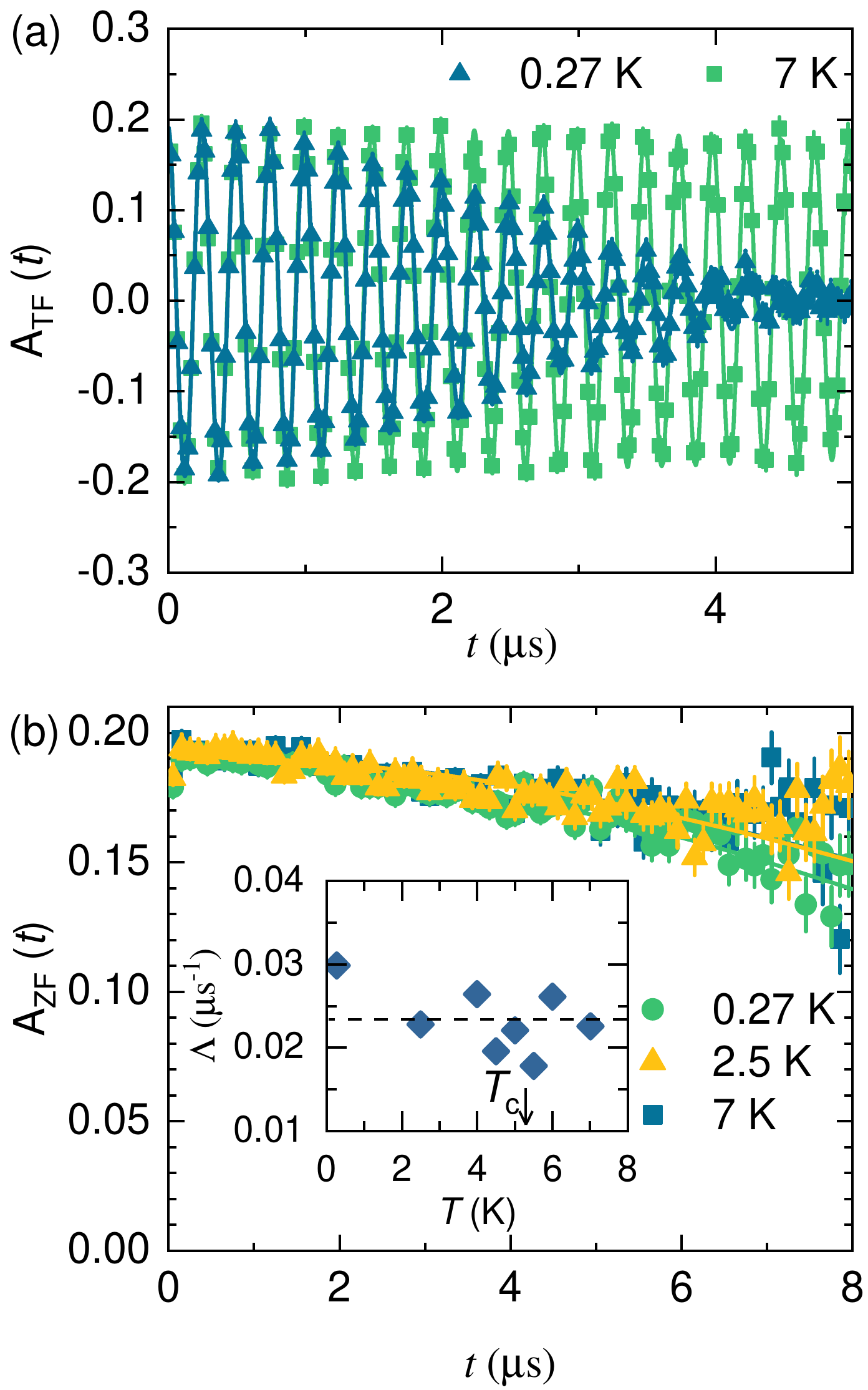}
\caption{(a) TF ${\mu}$SR time domain spectra obtained above and below $T_{\rm c}$ for \mbox{Ti$_4$Ir$_2$O} in an applied field of 30~mT (after field cooling the sample from above $T_{\rm c}$). (b) ZF $\mu$SR asymmetry spectra recorded at 0.27 and 7~K for Ti$_4$Ir$_2$O. Inset: temperature dependence of the electronic relaxation rate measured in zero magnetic field.}
\label{fig2}
\end{figure}

Ti$_4$Ir$_2$O crystallizes in the $\eta$-carbide type structure in the cubic space group $Fd\bar{3}m$, with a unit cell parameter of $a = 11.62931(2) $~\AA$ $ at room temperature \cite{Ma3}. In Figure \ref{fig1}(a), we show a schematic view of the Ti$_4$Ir$_2$O crystal structure along two orientations. The number of atoms in the unit cell is 112, of which 96 are metal atoms and 16 are oxygen. In the structure, titanium atoms occupy the $16c$ and the $48f$ Wyckoff positions, iridium atoms occupy the $32e$ Wyckoff positions, and oxygen atoms occupy the $16d$ Wyckoff positions, resulting in a formula of Ti$_{64}$Ir$_{32}$O$_{16}$ for one unit cell \cite{Ma3}. The prepared polycrystalline Ti$_4$Ir$_2$O samples were found to be single phase by means of powder X-ray diffraction measurement.

In Figure \ref{fig1}(b), we present high magnetic field (20~T) transport measurement of Ti$_4$Ir$_2$O down to 320 mK. The data was plot as the normalized resistivity $\rho$(H)/$\rho$(20 T) versus magnetic field $\mu_0 H$(T) at different constant below the critical temperature $T_{rm c}$. We can clearly observe the transition to the normal state from the superconducting state at high magnetic fields under all measured temperatures. Here, we determined the superconducting transition critical field $\mu_0 H$(T) at different temperatures as a midpoint of the transition. These high magnetic field transport measurement data together with those measured under lower fields were used to obtain the upper critical field $\mu_0H_{\rm c2}$(0) as shown in Figure \ref{fig1}(c).     

The upper critical field $\mu_0H_{\rm c2}$(0) of Ti$_4$Ir$_2$O was estimated using the Werthamer-Helfand-Hohenberg (WHH) approximation in the clean limit\cite{Ma3}:   

\begin{equation}
\mu_0 H_{c2} (T)= \frac{\mu_0 H_{c2} (0)}{0.73} h^\ast_{fit} (T/T_{\rm c}).
\label{eq:WHH}
\end{equation}
\begin{equation}
 h^\ast_{fit} (t) = (1-t) - C_{1}(1-t)^2 - C_{2}(1 - t)^4.
\end{equation}

where $t$ = $T/T_c$ and $T_c$ (5.30 K) is the transition temperature at zero magnetic field. In Figure \ref{fig1}(c), we show the $\mu_0H_{\rm c2}$(0) fitting of Ti$_4$Ir$_2$O using the Werthamer-Helfand-Hohenberg (WHH) approximation in the clean limit. We found the fitting line fairly well describes all the obtained experimental points, where $\mu_0H_{\rm c2}$(0) was determined to be 18.01 T. This value is exceeding by far than the Pauli paramagnetic limit $\mu_0 H_{\rm Pauli}$ of 9.85 T,  which is derived from $\mu_0 H_{\rm Pauli} \approx 1.86{\rm [T/K]} \cdot T_{\rm c}$ for a weak-coupling BCS superconductor. Here, this unusual high upper critical field indicates an unconventional superconductivity behavior in Ti$_4$Ir$_2$O.


\begin{figure*}[htb!]
\includegraphics[width=0.9\linewidth]{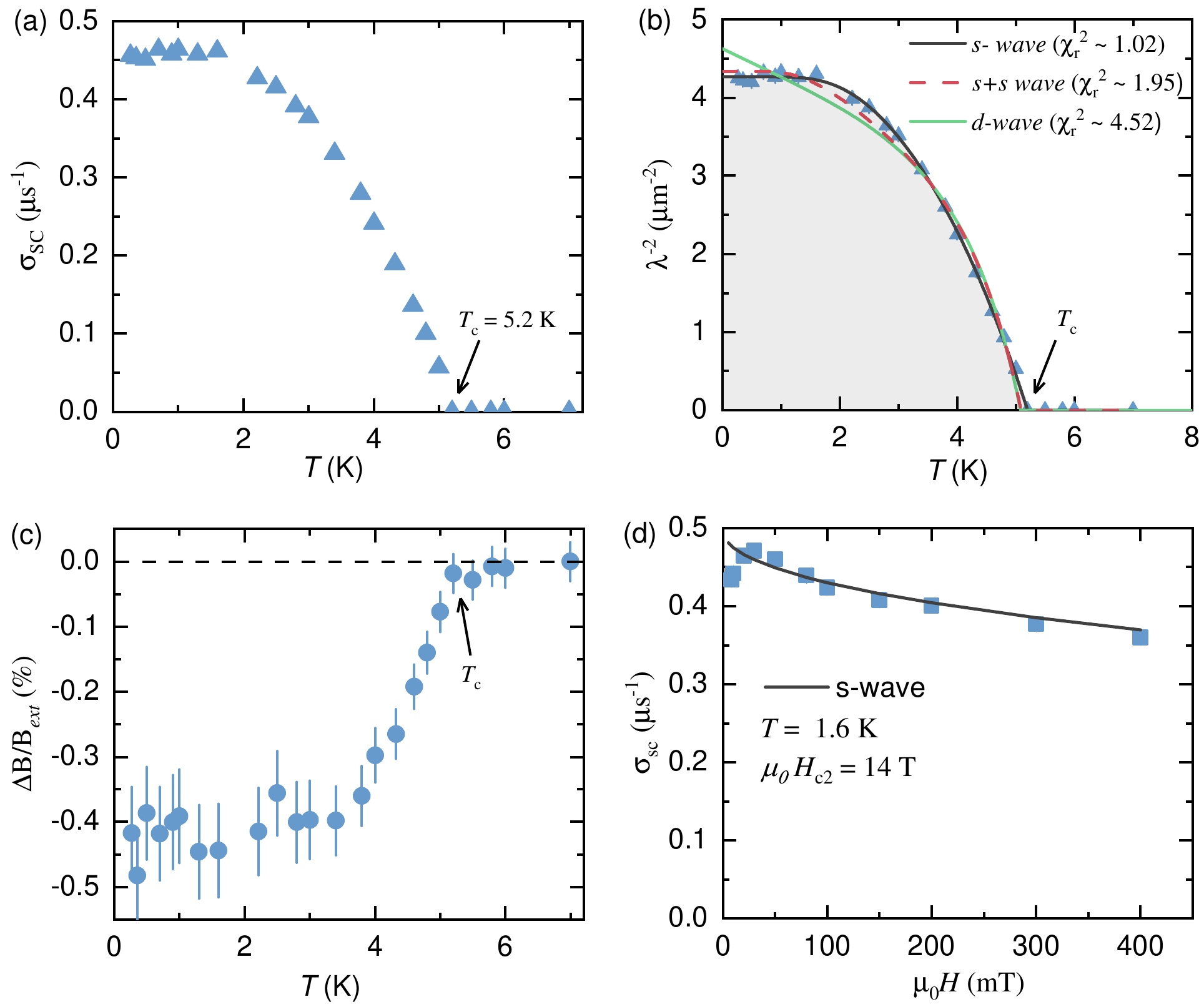}
\caption{(a) Temperature evolution of the superconducting muon spin depolarization rate ${\sigma}_{\rm sc}$ of Ti$_4$Ir$_2$O measured in an applied magnetic field of 30~mT. (b) Temperature evolution of ${\lambda}^{-2}(T)$  measured in an applied field ${\mu}_{\rm 0}H = 30~mT$. The solid and dashed lines represent fitting with different theoretical models as discussed in the text. For $s+s$-wave fitting, we fixed the gap values obtained from heat capacity data\cite{Ruan}.  
(c) Temperature dependence of the relative change of the internal field normalized to the external applied field, $\Delta B/B_{\rm ext}\left(= \frac {B_{\rm int}-B_{\rm ext}}{B_{\rm ext}}\right)$. 
(d) The field dependence of TF-relaxation rate ${\sigma}_{\rm sc}(B)$ measured at 1.6~K}
\label{fig3}
\end{figure*}

Figure~\ref{fig2}a shows TF-$\mu$SR spectra for Ti$_4$Ir$_2$O measured in an applied magnetic field of ~30~mT at temperatures above (7~K) and below (0.27~K) the critical temperature $T_{\rm c}$. A small relaxation, observed in TF-$\mu$SR spectra above  $T_{\rm c}$, can be attributed to the presence of random local fields associated with the nuclear magnetic moments. In the superconducting state, the formation of FLL creates an inhomogeneous distribution of magnetic field which leads to the increase of the relaxation rate of the $\mu$SR signal below $T_{\rm c}$.  

Magnetism, if present in the samples, may also enhance the muon depolarization rate and influence the interpretation
of the TF-$\mu$SR results. Therefore, we have carried out ZF-${\mu}$SR experiments above and below $T_{{\rm c}}$ to search for magnetism (static or fluctuating) in Ti$_4$Ir$_2$O. As seen from Fig.~\ref{fig2}b,we do not observe any noticeable difference in ZF-${\mu}$SR asymmetry spectra recorded at temperatures above and below $T_{{\rm c}}$. The ZF-${\mu}$SR asymmetry spectra can be well described by an exponential decay function, $A_{\rm ZF}(t) = A_0~G_{KT}~\exp(-\Lambda t)$ where $A_0$ is the initial asymmetry and $\Lambda$ is the depolarization rate. The inset of Fig.~\ref{fig2}b shows the temperature evolution of $\Lambda$, which shows no noticeable enhancement across $T_{{\rm c}}$. The maximum possible spontaneous flux density due to superconductivity can be estimated using \mbox{($\Lambda|_{0.27~\rm K}-\Lambda|_{7~\rm K})/(2\pi\gamma_{\mu}) = 1.49~\mu$T} which is several times smaller than that seen for well known TRS breaking superconductors \cite{LukeTRS, Hillier}. This demonstrates the absence of any spontaneous field in either the normal or the superconducting state of Ti$_4$Ir$_2$O. Therefore, this sample is non-magnetic and also the time-reversal symmetry is preserved in the superconducting state of this compound.

The absence of magnetism in Ti$_4$Ir$_2$O implies that the increase of the TF relaxation rate below $T_{\rm c}$ is attributed entirely to the flux-line lattice (FLL). Assuming a Gaussian field distribution, we analyzed the observed TF-$\mu$SR asymmetry spectra using the following functional form

\begin{equation}
A_{\rm TF}(t)=A_{0}\exp\left(\sigma^2t^2/2\right)\cos\left(\gamma_{\mu}B_{\rm int}t+\varphi\right)
\label{ATF}
\end{equation}

\noindent where $A_{0}$ refers to the initial asymmetry, $\gamma_\mu\simeq 851.615$~MHz/T is the muon gyromagnetic ratio, and ${\varphi}$ is the initial phase of the muon-spin ensemble, $B_{\rm int}$ corresponds to the internal magnetic field at the muon site, respectively and $\sigma$ is the total relaxation rate. $\sigma$ is correlated to the superconducting relaxation rate, $\sigma_{\rm SC}$, following the relation $\sigma=\sqrt{\sigma_{\rm nm}^2+\sigma_{\rm SC}^2}$ where $\sigma_{\rm nm}$ is the nuclear contributions that is assumed to be temperature independent. To estimate $\sigma_{\rm SC}$, we considered the value of $\sigma_{\rm nm}$ obtained above $T_{\rm c}$ where only nuclear magnetic moments contribute to the muon depolarization rate $\sigma$. The solid lines in Fig.~\ref{fig1}b depicts the fits to the observed spectra with Eq.~\ref{ATF}. 

In Figure~\ref{fig3}a, we have presented ${\sigma}_{\rm sc}$ as a function of temperature for Ti$_4$Ir$_2$O measured at an applied field of 30~mT. Below $T_{\rm c}$, the relaxation rate ${\sigma}_{\rm sc}$ increases from zero due to inhomogeneous field distribution caused by the formation of FLL, and saturates at low temperatures. In the following section, we show that the observed temperature dependence of ${\sigma}_{\rm sc}$, which reflects the topology of the superconducting gap, is consistent with the presence of the single gap on the Fermi surface of  Ti$_4$Ir$_2$O. Figure~\ref{fig3}c shows the temperature dependence of the relative change of the internal field normalized to the external applied field, $\Delta B/B_{\rm ext}\left(= \frac {B_{\rm \rm int}-B_{\rm ext}}{B_{\rm ext}}\right)$. As seen from the figure, internal field values in the superconducting state ($i.e. T<T_{\rm c}$) are lower than the applied field because of the diamagnetic shift, expected for type-II superconductors.

For a perfect triangular vortex lattice, ${\sigma}_{\rm sc}(T)$ is directly related to the London magnetic penetration depth ${\lambda}(T)$ by~\cite{Brandt,Brandt2}:
\begin{equation}
\frac{\sigma_{\rm sc}^2(T)}{\gamma_\mu^2}=0.00371\frac{\Phi_0^2}{\lambda^4(T)}.
\label{Sigma}
\end{equation}
\noindent where, $\Phi_0$ = 2.068~$\times$~10$^{-15}$~Wb is the magnetic flux quantum. It is important to note that Eq.~\ref{Sigma} is valid when the separation between the vortices is smaller than ${\lambda}$~\cite{Brandt}. We have analyzed the temperature dependence of the magnetic penetration depth, $\lambda^{-2}(T)$ to unveil the superconducting gap structure of Ti$_4$Ir$_2$O.

Within the London approximation ($\lambda \gg {\xi}$), ${\lambda}(T)$ can be described by the following expression,~\cite{Suter,Tinkham, Prozorov}
\begin{equation}
\frac{\lambda^{-2}(T,\Delta_{0,i})}{\lambda^{-2}(0,\Delta_{0,i})}=
1+\frac{1}{\pi}\int_{0}^{2\pi}\int_{\Delta(_{T,\varphi})}^{\infty}\left(\frac{\partial f}{\partial E}\right)\frac{EdE}{\sqrt{E^2-\Delta_i(T,\varphi)^2}},
\label{Lambda}
\end{equation}
\noindent where $f=\left[1+\exp\left(E/k_{\rm B}T\right)\right]^{-1}$ is the Fermi function, $\varphi$ is the azimuthal angle  along the Fermi surface and ${\Delta}_{0,i}\left(T\right)={\Delta}_{0,i}~{\Gamma}\left(T/T_{\rm c}\right)$~g($\varphi$). ${\Delta}_{0,i}$ is the maximum gap value at $T=0$~K. The temperature dependence of the gap is described by the expression \mbox {${\Gamma}\left(T/T_{\rm c}\right)=\tanh\left\{1.82\left[1.018\left(T_{\rm c}/T-1\right)\right]^{0.51}\right\}$}~\cite{carrington}. For $s$-wave gap and a nodal $d$-wave, the angular dependence g($\varphi$) corresponds to 1 and $\mid cos(2\varphi)\mid$ respectively. We have considered three models in our analysis: (i) an isotropic $s$-wave gap, (ii) a combination of two $s$-wave gaps of different size, and (iii) a nodal $d$ wave model. 

As seen from the Figure~\ref{fig3}b, the experimentally obtained ${\lambda}^{-2}(T)$ dependence can be best described using a single $s$-wave model yielding a gap value of $\Delta_0$ = 0.92(8)~meV and $T_{\rm c}$ = 5.19(3)~K. Previously from heat capacity analysis,  Ruan $et~al.$\cite{Ruan} predicted the presence of two superconducting gaps ($\Delta_{0,1}$ = 1.37~meV and $\Delta_{0,2}$ = 0.57~meV). Therefore, to test this possible multi-gap scenario, we used these gap values and kept them fixed in our analysis. 

For the two-gap scenario, we have used the weighted sum of two gaps:
\begin{equation}
\frac{\lambda\textsuperscript{-2}(T)}{\lambda\textsuperscript{-2}(0)}= x\frac{\lambda\textsuperscript{-2}(T,\Delta_{0,1})}{\lambda\textsuperscript{-2}(0,\Delta_{0,1})}+(1-x)\frac{\lambda\textsuperscript{-2}(T,\Delta_{0,2})}{\lambda\textsuperscript{-2}(0,\Delta_{0,2})}.
\end{equation}
\noindent Here $x$ is the weight associated with the larger gap and $\Delta_{0,i}$ ($i$ = 1, 2 are the gap indices) are the gaps.

A close look at the goodness of fitting suggests that the single $s$-wave model gives the lowest value of $\chi_r^{2}$ indicating the best fit to the observed data. We note that while fitting the experimental data keeping two gaps as free parameters, the weight of the higher gap ($x$) was reaching a value close to 0, implying the single gap scenario is more appropriate. Also $d$-wave gap symmetry was also tested, but were found to be inconsistent with the data (for example, see the green solid line in Fig~3b). Thus, from the $\mu$SR experiments, we confirm the presence of a single fully-gapped superconducting state in Ti$_4$Ir$_2$O. 

We also investigated the field dependence of TF-relaxation rate ${\sigma}_{\rm sc}(B)$ measured at 1.6~K (see Fig~\ref{fig3}d). For these measurements, each point was obtained by field-cooling the sample from 10~K (above $T_{\rm c}$) to 1.6~K.  ${\sigma}_{\rm sc}$ first increases to a maximum with increasing magnetic field followed by a continuous decrease up to the highest field (400~mT) studied as expected for a single gap superconductor with an ideal triangular vortex lattice. Interestingly, the observed ${\sigma}_{\rm sc}(B)$ curve at fields above the maximum, can be well modelled using the Brandt formula (for a single $s$-wave gap superconductor)~\cite{Brandt2} with an upper critical field $\mu_0 H_{\rm c2}(0.27~K)= 14$~T at 1.6~K.

While the temperature- and field-dependence of the London penetration depth $\lambda_{eff}^{-2}$ and henceforth of the superfluid density can be well understood as a fully gapped s-wave superconductor, the $T_{\rm c}$/$\lambda_{eff}^{-2}$ ratio for Ti$_4$Ir$_2$O is comparable to those of high-temperature unconventional superconductors. In the context of the Bose-Einstein Condensation (BEC) to BCS crossover \cite{Uemura1}, systems with a small ratio of $T_{\rm c}$/$\lambda_{eff}^{-2}$ (approximately 0.00025-0.015) are categorized on the BCS-like side, indicative of conventional superconductivity. In contrast, a large ratio in the range of 1-20, along with a linear relationship between $T_{\rm c}$ and $\lambda_{eff}^{-2}$, is typically observed on the BEC-like side, signifying unconventional superconductivity. This framework has been historically utilized to differentiate between BCS-like (conventional) and BEC-like (unconventional) superconductors. The results obtained for Ti$_4$Ir$_2$O show a ratio of $T_{\rm c}$/$\lambda_{eff}^{-2}$ ${\simeq}$ 1.22, closely aligning with the ratio of approximately 1 seen in electron-doped cuprates. This finding, alongside previously reported high upper critical fields and thermodynamic evidence of an FFLO state, provides strong support for an unconventional pairing mechanism in Ti$_4$Ir$_2$O.

In conclusion, we provide the first microscopic investigation of superconductivity in the $\eta$-carbide-type suboxide
Ti$_4$Ir$_2$O with a bulk probe. Namely, the zero-temperature magnetic penetration depth ${\lambda}_{eff}\left(0\right)$ and the temperature as well as the field dependence of ${\lambda_{eff}^{-2}}$ were studied by means of ${\mu}$SR experiments. 

We have demonstrated the isotropic fully gap pairing and the preserved time-reversal symmetry in the SC state of this material. Interestingly, the $T_{\rm c}$/$\lambda_{eff}^{-2}$ ratio is comparable to those of high-temperature unconventional superconductors, pointing to the unconventional nature of superconductivity in Ti$_4$Ir$_2$O. This result -- together with earlier findings of the high upper critical field larger than the Pauli limit and the thermodynamic signatures for an FFLO state -- hints towards an unconventional pairing mechanism in this material. These results identify Ti$_4$Ir$_2$O as time-reversal-invariant fully gapped unconventional superconductor.

\begin{acknowledgments}
The muon spectroscopy studies were performed at the Swiss Muon Source (S${\mu}$S) Paul Scherrer Insitute, Villigen, Switzerland. This work was supported by the Swiss National Science Foundation under Grant No. PCEFP2\_194183. We acknowledge Dr. Christopher Baines for the technical support provided during the experiment. Z.G. acknowledges support from the Swiss National Science Foundation (SNSF) through SNSF Starting Grant (No. TMSGI2${\_}$211750). A portion of this work was performed at the National High Magnetic Field Laboratory, which is supported by National Science Foundation Cooperative Agreement No. DMR-1644779 and the State of Florida."
\end{acknowledgments}

\end{document}